\documentclass[final,3p,times,twocolumn]{elsarticle}
 \biboptions{comma,sort&compress}
\usepackage{here}
 \usepackage{graphicx}
  \usepackage{epsfig}
\usepackage{slashed}
\usepackage{amssymb,amsmath}

\def\nin{\noindent}
\def\beq{\begin{equation}}
\def\eeq{\end{equation}}
\def\bea{\begin{eqnarray}}
\def\eea{\end{eqnarray}}

\journal{Nuc. Phys. (Proc. Suppl.)}

\begin{document}

\begin{frontmatter}



\title{Low energy limit of QCD and the emerging of confinement}

 \author[label1]{Marco Frasca\corref{cor1}}
  \address[label1]{Via Erasmo Gattamelata, 3\\
00176 Rome (Italy).}
\cortext[cor1]{Speaker}
\ead{marcofrasca@mclink.it}



\begin{abstract}
\noindent
After a derivation of low-energy limit of QCD, being this a non-local Nambu-Jona-Lasinio model, we are able to show that confinement emerges as a two-loop correction to the gluon propagator. One-gluon exchange is not enough as recently shown in literature about studies on the gluon propagator in the Landau gauge.

\end{abstract}


\end{frontmatter}


\section{Introduction}
\nin

Quantum field theory can be approached with analytical techniques if a set of classical solutions to field theories are known. This can be understood when we look at standard techniques like small perturbation theory where one generally starts from classical solutions of the free theory. Here we show that such a set of solutions exists for a class of nonlinear field theories giving us the chance to analyze their behavior in the infrared limit. One of these is Yang-Mills theory.

When the low-energy behavior of Yang-Mills theory is known one can approach quantum chromodynamics (QCD) in the same limit. Things stay this way as a knowledge of the propagator of the gluon can yield the proper low-energy limit of QCD\cite{Frasca:2011bd, Frasca:2008zp}. In this way, a non-local Nambu-Jona-Lasinio model is recovered that has already discussed in \cite{Hell:2008cc}. But, in our case, the form factor is completely fixed by QCD rather than phenomenologically guessed. Such a form factor agrees in a strikingly good way with the one obtained for an instanton liquid \cite{Schafer:1996wv}. This gives a sound evidence that instantons could play a relevant role for the ground state of the theory. Finally, it is essential to point out that another approach provides a similar link between QCD and Nambu-Jona-Lasinio model \cite{Kondo:2010ts}.

We will show in this paper how confinement can emerge in such a scenario, confirming a recent finding in literature that one-gluon exchange is not enough to grant it \cite{Gonzalez:2012hx}, given a proper form for the gluon propagator in the infrared limit and higher loops are considered.

\section{Classical fields}

The simplest theory to analyze, starting from a classical standpoint, is the following
\begin{equation}
\label{eq:cphi}
   \Box\phi+\lambda\phi^3=j.
\end{equation}
When $j=0$ an exact solution exists and is $\phi = \mu\left(2/\lambda\right)^\frac{1}{4}{\rm sn}(p\cdot x+\theta,i)$
being {\rm sn} an elliptic Jacobi function and $\mu$ and $\theta$ two integration constants. This solution holds provided the dispersion relation $p^2=\mu^2\sqrt{\lambda/2}$ holds. This appears to be a free massive solution notwithstanding we started from a massless theory. Mass originates from the nonlinear term but the coupling must be taken finite as a small perturbation expansion fails to recover it. We want to solve eq. (\ref{eq:cphi}) when $\lambda\rightarrow\infty$. In order to achieve this aim we consider a technique firstly devised in the '80s \cite{Cahill:1985mh}. One considers $\phi$ a functional of the current $j$ and makes a power series of it. It is not difficult to realize that this argument is consistent when
\begin{equation}
\label{eq:ps}
\begin{split}
   \phi(x)&=\mu\int d^4x'\Delta(x-x')j(x') \nonumber \\
   &+\mu^2\lambda\int d^4x'd^4x''\Delta(x-x')[\Delta(x'-x'')]^3j(x')+O(j(x)^3)
\end{split}
\end{equation}
being $\Delta(x-x')$ a solution to the nonlinear equation $\Box\Delta(x-x')+\lambda[\Delta(x-x')]^3=\mu^{-1}\delta^4(x-x')$. Such a series has an interesting leading term. This term, when introduced in a quantum field theory, will yield a Gaussiam term in the generating functional that is the hallmark of a trivial theory. We realize that this set of classical solutions are peculiar to a trivial theory in a leading order. This means that such a theory is manageable notwithstanding the presence of strong nonlinearities. But this program can only be complete if we are able to get the Green function of the theory. In this case this can be easily obtained by starting from the theory at $d=1+0$. So, the equation to be solved is $\partial_t^2\Delta_0(t-t')+\lambda[\Delta_0(t-t')]^3=\mu^2\delta(t-t')$ and this is easily obtained \cite{Frasca:2005sx}. A Lorentz boost or a gradient expansion provides the complete solution
\begin{equation}
\label{eq:green}
   \Delta(p)=\sum_{n=0}^\infty(2n+1)\frac{\pi^2}{K^2(i)}\frac{(-1)^{n}e^{-(n+\frac{1}{2})\pi}}{1+e^{-(2n+1)\pi}}
   \frac{1}{p^2-m_n^2+i\epsilon}
\end{equation}
being $m_n=(2n+1)(\pi/2K(i))\left(\lambda/2\right)^{\frac{1}{4}}\mu$ and $K(i)\approx 1.3111028777$ an elliptic integral. This Green function appears to be consistent with a strong coupling expansion.

The approach of current expansion devised in the '80s \cite{Cahill:1985mh} was conceived to manage QCD. The leading order for the Yang-Mills part of the theory was Gaussian. So, we try to see if, assuming to exist an infrared trivial fixed point for Yang-Mills theory, a set of classical solutions for Yang-Mills equations exists to support this approach as already happens for the scalar field. We start from the equations of motion
\begin{equation}
\begin{split}
\partial^\mu\partial_\mu A^a_\nu&-\left(1-\frac{1}{\xi}\right)\partial_\nu(\partial^\mu A^a_\mu) 
+gf^{abc}A^{b\mu}(\partial_\mu A^c_\nu-\partial_\nu A^c_\mu)\nonumber \\
&+gf^{abc}\partial^\mu(A^b_\mu A^c_\nu) \\
&+g^2f^{abc}f^{cde}A^{b\mu}A^d_\mu A^e_\nu = -j^a_\nu.
\end{split}
\end{equation}
In this case, we note that the homogeneous equations can be solved by looking at instanton like solutions $A_\mu^a(x)=\eta_\mu^a\phi(x)$ being $\eta_\mu^a$ a set of constants. The problem here is that we have to fix the gauge. We consider the case of the Lorenz gauge that is equivalent to the Landau gauge in quantum field theory. In this case the equation collapses to $\partial^\mu\partial_\mu\phi+Ng^2\phi^3=-j_\phi$ and our problem is equivalent to the one of the scalar field we discussed before (we just note that this is not true for other gauge choices where we can only claim an asymptotic correspondence \cite{Frasca:2009yp}). Then, we are able to write down immediately the gluon propagator in the Landau gauge, given eq.(\ref{eq:green}), just setting $\lambda=Ng^2$ and introducing the factor $\delta_{ab}\left(\eta_{\mu\nu}-p_\mu p_\nu/p^2\right)$. This conclusion implies that instantons are relevant in the ground state of the theory.

\begin{figure}[hbt] 
\centerline{\includegraphics[width=6.cm]{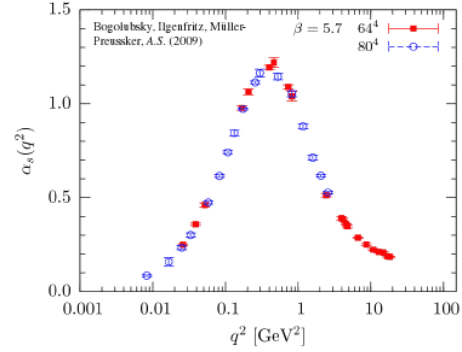}}
\caption{\scriptsize Running coupling for $64^4$ and $80^4$ at $\beta=5.7$ \cite{Bogolubsky:2009dc}.}
\label{fig:rc} 
\end{figure}

Our fundamental starting hypothesis, that Yang-Mills theory admits an infrared trivial fixed point, should be tested on the lattice to get support. This is indeed the case. From the German group \cite{Bogolubsky:2009dc} the running coupling is seen to go to zero as momenta lower (see fig. \ref{fig:rc}). 
\nin
Similarly, the French group obtained such a behavior for the running coupling with a different definition \cite{Boucaud:2002fx}. They show a perfect consistency with an instanton liquid model in agreement with the scenario we are depicting here.

\section{Quantum fields}

\begin{figure}[hbt] 
\centerline{\includegraphics[width=8.cm]{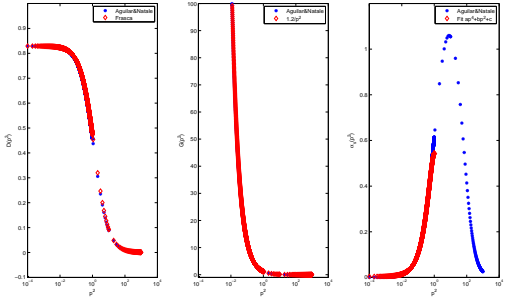}}
\caption{\scriptsize Comparison of our propagators with numerical solution of Dyson-Schwinger equations as in \cite{Aguilar:2004sw}. Running coupling is given on the right.}
\label{fig:an} 
\end{figure}

Quantum theory for the scalar field is managed with the generating functional $Z[j]={\cal N}\int[d\phi]\exp\left\{i\int d^4x\left[(\partial\phi)^2-\lambda\phi^4/4\right]\right\}$. We rescale the space-time coordinates as $x\rightarrow\sqrt{\lambda}x$ and take a strong coupling expansion $\phi=\sum_{n=0}^\infty\lambda^{-n}\phi_n$. Then, we see that the leading order reduces to solve the equation $\Box\phi_0+\lambda\phi_0^3=j$ that we already know how to manage. This means that the leading order is just a Gaussian functional with the propagator given by eq.(\ref{eq:green}) and the current expansion discussed above \cite{Frasca:2010ce}. Next-to-leading order can be also computed. Our fundamental result is that the massless scalar field theory in four dimensions is infrared trivial \cite{Frasca:2010ce}. Mass spectrum is given by $m_n=(2n+1)(\pi/2K(i))\left(\lambda/2\right)^\frac{1}{4}\mu$ that  represents free particles with a superimposed spectrum of a harmonic oscillator. For the Yang-Mills generating functional a similar Gaussian functional is obtained
\begin{equation*}
     Z_0[j]=N\exp\left[\frac{i}{2}\int d^4x'd^4x''j^{a\mu}(x')D_{\mu\nu}^{ab}(x'-x'')j^{b\nu}(x'')\right].
\end{equation*}
given the current expansion $A_\mu^a=\Lambda\int d^4x' D_{\mu\nu}^{ab}(x-x')j^{b\nu}(x')+O\left(1/\sqrt{N}g\right)+O(j^3)$ and the propagator $D_{\mu\nu}^{ab}(p)=\delta_{ab}\left(\eta_{\mu\nu}-\frac{p_\mu p_\nu}{p^2}\right)\Delta(p)$ being $\Delta(p)$ given by eq.(\ref{eq:green}) with $\lambda=Ng^2$. The spectrum is given by free glue particles with a superimposed spectrum of a harmonic oscillator. Similarly, for the ghost field, our instanton solutions decouple it from the gauge field yielding a free particle propagator for a massless field. These properties of the quantum Yang-Mills field represent the so-called ``decoupling solution'' \cite{Aguilar:2004sw,Boucaud:2006if,Frasca:2007uz}. This solution is the one recovered in lattice computations \cite{Bogolubsky:2007ud,Cucchieri:2007md,Oliveira:2007px}. Numerical solution of Dyson-Schwinger equations as given in \cite{Aguilar:2004sw} (see fig.\ref{fig:an}) are in perfect agreement with it.
\nin

\begin{figure}[hbt] 
\centerline{\includegraphics[width=6.cm]{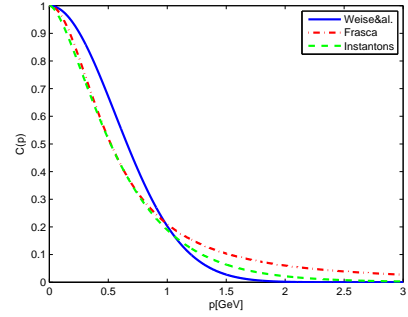}}
\caption{\scriptsize Comparison of our form factor with those provided in Hell et al.\cite{Hell:2008cc} and instanton liquid \cite{Schafer:1996wv} as given in \cite{Frasca:2011bd}.}
\label{fig:ff} 
\end{figure}

Once we know the behavior of Yang-Mills theory, we are able to get the low-energy behavior of QCD. One can show that it takes the form \cite{Frasca:2011bd}
\begin{equation}
      S=\int d^4x\left[\frac{1}{2}(\partial\sigma)^2-\frac{1}{2}m_0^2\sigma^2\right]+S_q
\end{equation}
where the $\sigma$ field arises from the gluon propagator in the Gaussian generating functional of the Yang-Mills action and is the contribution from the mass gap of the theory, being $m_0=(\pi/2K(i))\sqrt{\tilde\sigma}$ and $\tilde\sigma$ is the string tension ($\approx (440\ MeV)^2$). For the quark fields one gets
\begin{equation}
\begin{split}
      S_q&=\sum_q\int d^4x\bar q(x)\left[i{\slashed\partial}-m_q\right]q(x) \\  
     &-g^2\int d^4x'\Delta(x-x')\times \\
     &\sum_q\sum_{q'}\bar q(x)\frac{\lambda^a}{2}\gamma^\mu\bar q'(x')\frac{\lambda^a}{2}\gamma_\mu q'(x')q(x) \\
      &+O\left(\frac{1}{\sqrt{N}g}\right)+O\left(j^3\right). 
\end{split}
\end{equation}
This result recovers the non-local Nambu-Jona-Lasinio model given in \cite{Hell:2008cc} but directly from QCD provided the form factor is
\begin{equation}
      {\cal G}(p)=-\frac{1}{2}g^2\Delta(p)=\frac{G}{2}{\cal C}(p)
\end{equation}
being $\Delta(p)$ that in eq.(\ref{eq:green}), ${\cal C}(0)=1$ and $2{\cal G}(0)=G$ the well-known Nambu-Jona-Lasinio coupling. So, the value of $G$ is fixed by the gluon propagator. We can compare this form factor both with the one guessed in \cite{Hell:2008cc} and the one from an instanton liquid \cite{Schafer:1996wv} in fig. \ref{fig:ff}. The result is strikingly good for the latter showing how consistently our technique represents Yang-Mills theory through instantons.
\nin
Finally, the $\sigma$ field introduced by the gluon field contributes to the $\sigma$ field arising after bosonization in the Nambu-Jona-Lasinio model \cite{Ebert:1994mf} giving a renormalized coupling constant $G_{eff}=1/(1/G+m_0^2)$. This field is effectively interacting with the other fields when these are given as bound states. This field can be identified with the $\sigma$ or f0(600) resonance, whose mass has been estimated to $m_\sigma=441^{+16}_{-8}\ MeV$ \cite{Caprini:2005zr} and $457^{+14}_{-13}\ MeV$ \cite{GarciaMartin:2011jx}. Our estimation, in the Nambu-Jona-Lasinio contact approximation that provides $m^*=214\ MeV$ for quark effective mass and $m_\pi=139.7\ MeV$, is $m_\sigma=\sqrt{4m^{*2}+m_\pi^2}=451\pm 20\ MeV$. The error should be ascribed to the string tension that is estimated with an error of $20\ MeV$. Notwithstanding the rough approximation, this model provides a very good value for this mass.

\section{Confinement}

In order to verify if, at leading order (one-gluon exchange), the scenario we depicted yields a confining theory we need to evaluate the integral for the interquark potential \cite{Gonzalez:2012hx}
\begin{equation}
   V({\bf x})=-\frac{2C_F}{\pi}\int d^3p\alpha_s({\bf p})\Delta({\bf p})e^{i{\bf p}\cdot{\bf x}}.
\end{equation}
Here, one has to use the running coupling $\alpha_s({\bf p})$ that, in the infrared limit, we have proved to go like $p^4$ \cite{Boucaud:2002fx,Frasca:2010ce}. This implies for the potential
\begin{equation}
   V(r)=-\frac{\alpha_s(0)}{\Lambda^4}\frac{1}{r}\frac{\partial^4}{\partial r^4}
   \sum_{n=0}^\infty(2n+1)\frac{\pi^2}{K^2(i)}\frac{(-1)^{n}e^{-(n+\frac{1}{2})\pi}}{1+e^{-(2n+1)\pi}}e^{-m_nr}   
\end{equation}
and, due to massive excitations, one gets a screened potential. This appears to agree very well with the conclusions given in \cite{Gonzalez:2012hx} but not in agreement with Cornell potential observed on the lattice for quenched simulations. So, if one just stops at the decoupling solution in the deep infrared, confinement is not recovered. This can be obtained if one is able to compute the next-to-leading order correction to the decoupling solution. For our scenario, this is presented in \cite{Frasca:2008gi} yielding a two-loop sunrise diagram. The correction to the leading order propagator takes the form
\begin{equation}
\begin{split}
    F(\lambda,p^2)&=\sum_{n_1,n_2,n_3}\int\frac{d^4p_1}{(2\pi)^4}\frac{d^4p_2}{(2\pi)^4}
    \frac{B_{n_1}}{p_1^2-m_{n_1}^2}
    \frac{B_{n_2}}{p_2^2-m_{n_2}^2}\times \nonumber \\
    &\frac{B_{n_3}}{(p-p_1-p_2)^2-m_{n_1}^2}
\end{split}
\end{equation}
with $m_{n_i}=(2n+1)(\pi/2K(i))\left(\lambda/2\right)^{\frac{1}{4}}\Lambda$, $\lambda=Ng^2$ and $B_{n_i}$ can be obtained from eq.(\ref{eq:green}). Techniques to evaluate this integral at small momenta, as we need, are well-known \cite{Caffo:1998du} and the result can be stated into the form
\begin{equation}
   \Delta_{\rm 2L}(p)=
   \Delta(p)\left[1+\frac{c_1}{\lambda\frac{1}{2}}+\frac{1}{\lambda}\left(c_2-c_3\frac{p^2}{\Lambda^2}\right)
   +O\left(\lambda^{-\frac{3}{2}}\right)\right].
\end{equation}
being $c_i$ some inessential numerical constants. This correction provides the needed $p^4$ Gribov contribution to the propagator to get a linear term in the potential, taking into account the behavior of the running coupling that increases as the fourth power of momenta. Our scenario is in striking agreement with the numerical solution of truncated Dyson-Schwinger equations as shown in fig. \ref{fig:an}. Truncated Dyson-Schwinger equations where considered in \cite{Gonzalez:2012hx} and our conclusions agree at one-gluon exchange level as they should.

We note that a regularization scheme must enter into this computation but our conclusions depend explicitly from a cut-off $\Lambda$. Firstly, we point out the the sums on $n_i$ remove some singular behavior being zero in such cases and the explicit physical cut-off $\Lambda$, in the infrared limit, is strictly linked to the string tension that is obtained from experiments. So, the argument for confinement appears consistent at this stage.

\section{Conclusions}
\nin
We provided a strong coupling expansion both for classical and quantum field theory of a massless quartic scalar field and pure Yang-Mills theory. A set of classical solutions is proved to exist for the Yang-Mills field, instantons, that support the view of a trivial infrared fixed point. A low-energy limit of QCD is so obtained that reduces to a non-local Nambu-Jona-Lasinio model with all the parameters and the form factor properly fixed by QCD. $\sigma$ meson gets a value of mass in close agreement with recent studies on $\pi\pi$-scattering and its very nature is mostly gluonic. Confinement can emerge in all this picture at two-loop level. Overall agreement with numerical studies of truncated Dyson-Schwinger equations and lattice studies make this scenario really compelling.

\section*{Acknowledgements}
\nin
I would like to thank Marco Ruggieri for useful comments and the numerical solution of Dyson-Schwinger equations.








\end{document}